# Teasing out the overall survival benefit with adjustment for treatment switching to other therapies

Yuqing Xu[1], Meijing Wu[2], Weili He[2], Qiming Liao[3], Yabing Mai[2]

**Abstract**: In oncology clinical trials, characterizing the long-term overall survival (OS) benefit for an experimental drug or treatment regimen (experimental group) is often unobservable if some patients in the control group switch to drugs in the experimental group and/or other cancer treatments after disease progression. A key question often raised by payers and reimbursement agencies is how to estimate the true benefit of the experimental drug group on overall survival that would have been estimated if there were no treatment switches. Several commonly used statistical methods are available to estimate overall survival benefit while adjusting for treatment switching, ranging from naive exclusion or censoring approaches to more advanced methods including inverse probability of censoring weighting (IPCW), iterative parameter estimation (IPE) algorithm or rank-preserving structural failure time models (RPSFTM). However, many clinical trials now have patients switching to different treatment regimens other than the test drugs, and the existing methods cannot handle more complicated scenarios. To address this challenge, we propose two additional methods: stratified RPSFTM and random-forest-based prediction. A simulation study is conducted to assess the properties of the existing methods along with the two newly proposed approaches.

[1] Department of Statistics, University of Wisconsin-Madison, Madison, WI, USA
[2] AbbVie Inc. Data Science and Statistics, North Chicago, IL, USA
[3] Department of Statistics, ViiV Healthcare, Raleigh-Durham, NC, USA

**Correspondence** Yuqing Xu, Department of Statistics, University of Wisconsin-Madison, Madison, WI 53705, USA
Email: yxu253@wisc.edu; emmayuqingxu@gmail.com

# 1 Introduction

In randomized clinical trial setting, treatment switching is defined as patients in the control group switching to the experimental drug and/or other cancer treatments. It is common in oncology clinical trials on cancer treatments and also occurs in trials of treatments for other diseases. If the switched therapy is effective, an intention-to-treat (ITT) analysis approach, which is often used in such trials, will underestimate the actual long-term overall survival (OS) benefit of the test drug had there been no switch. Therefore, if the clinical objective is to compare the long-term effectiveness of an experimental treatment with placebo or an active comparator, the ITT analysis often provides a biased assessment. A key question often asked by payers and reimbursement agencies is on how to estimate the true benefit of an experimental drug on overall survival that would have been observed if there were no treatment switches.

Several commonly used statistical methods are available to estimate overall survival benefit while adjusting for treatment switching, ranging from naive exclusion or censoring approaches to more advanced methods including inverse probability of censoring weighting (IPCW), iterative parameter estimation (IPE) algorithm or rank-preserving structural failure time models (RPSFTM). Naïve methods often provide largely biased results, while RPSFTM and IPE can only adjust for switches to the experimental treatment. They cannot handle cases where switches occur to multiple levels of treatments. Furthermore, IPCW is not very stable and its confidence intervals are wide when sample sizes are small or proportion of switchers is relatively large. However, more and more clinical trials now have patients switching to multiple treatments other than the test drug, and the existing methods are limited in handling this more complicated scenario. Therefore, new methods are needed to further reduce potential biases and enhance the robustness of estimation and tests in assessing overall survival with the presence of treatment switching, especially when there are patients switching to multiple treatments. In this manuscript, we propose two new methods - Stratified RPSFTM and random-forest-based prediction, to overcome these difficulties.

In section 2, we briefly introduce the current methods on treatment switching from control arm to experimental treatment, including simple methods such as ITT, excluding switchers and censoring switchers, and advanced methods, such as IPCW, RPSFTM and IPE. In section 3, we describe in detail the two newly proposed methods for multi-level treatment switching: Stratified RPSFTM and random-forest-based prediction. Section 4 includes a simulation study to evaluate the operating characteristics of the existing methods and the newly proposed methods in analysis of overall survival benefit of a treatment in presence of multi-level treatment switching under different scenarios. We conclude with a discussion in Section 5.

## 2 Current Methods

### 2.1 Simple methods

The ITT analysis is to evaluate overall survival benefit by comparing the observed overall survival time between randomized control arm and experimental arm. ITT approach serves as a reference for other sophisticated approaches. However, with presence of treatment switching, if patients benefit from another active therapy, the true overall survival benefit associated with the original experimental treatment may be inaccurately estimated using the ITT approach, and this in turn can affect the subsequent cost effectiveness analyses that rely on overall survival evidence.

Other simple methods include excluding switchers and censoring switchers at the time of switching. Excluding switchers breaks randomization, reduces the sample size in control arm, and often leads to bias. Censoring switchers at the time of switch also brings bias to analysis as the fundamental assumption that censoring is independent of survival time is unlikely to hold in this situation.

### 2.2 Advanced methods

In order to reduce the potential biases in evaluating overall survival benefit of a treatment in the presence of treatment switching, the following advanced methods that are currently available from literature are introduced here.

#### 2.2.1 Inverse Probability Censoring Weighting (IPCW)

Robins and Finkelstein (2000)[1] proposed IPCW as a treatment effect estimation method in the presence of dependent censoring used in a marginal structural model (MSM)[2]. For treatment switching cases, patients are artificially censored at the time of switch. Weights are increased for those patients who did not switch but have similar baseline characteristics to patients who did switch. More specifically, the weights will be obtained for each patient *i* based on the inverse probability of patient *i* remaining in the control treatment until time *t*. Then treatment effect will be estimated using weighted survival analysis methods, e.g. weighted Cox regression model or weighted Kaplan-Meier curve[3].

The assumption of the method is "no unmeasured confounders" - all baseline covariates and all post-baseline time dependent confounders that predict both treatment switch and outcome are included[4]. However, this assumption is unlikely to hold and cannot be tested using the observed data. Some key predictors of treatment switching are usually not collected in RCTs. In addition, the IPCW method cannot work if there are any covariates that directly determine treatment switches or not (that is, the probability equals 1). Hence IPCW is less stable and its confidence intervals are wide when the sample sizes are small or proportion of switchers is relatively large.

### 2.2.2 Rank Preserving structure failure time model (RPSFTM)

The RPSFTM method, proposed by Robins and Tsiatis (1991)[5], is the structural or strong version of the "accelerated failure time model with time-dependent covariates" of Cox and Oakes (1984)[6]. The RPSFTM method uses a counterfactual framework to estimate the treatment effect, where counterfactual survival time refers to time that would have been observed if no experimental treatment were received. The method splits the observed event time $T_i$ for each patient into two parts: the event time when the patient is on the control treatment $T_{i,CON}$, and the event time when the patient is on the intervention treatment, $T_{i,TRT}$. The RPSFTM method relates $T_i$ to the counterfactual event time $U_i$ using accelerated failure time (AFT) model with an acceleration factor $\exp(-\psi)$:

$$U_i = T_{i,CON} + e^{\psi} T_{i,TRT}$$

Due to randomization, assume $U_i$ distribution is the same for both arms. Then for a range of ψ values, calculate $U_i$ and test if $U_i$ is equal for all patients (using a standard survival analysis test statistic Z). Select ψ that best satisfies test statistic Z (ψ) = 0. This estimation is the so-called g-estimation proposed by Robins et al[7].

One concern of applying RPSFTM is that survival time cannot be reduced in a similar fashion for censored patients without event. This would introduce potential bias due to informative censoring. Hence it is suggested in White et al[8] and some other literature that when modeling counterfactual survival time $U_i$, one can re-censor all patients at

$$U_i^* = \min(U_i, C_i, e^{\psi} C_i)$$

The key assumption of RPSFTM is the "common treatment effect", which means that the untreated outcome U is independent of the randomized groups. Previous simulation results of Latimer N, Abrams K, Lambert P, et al. showed that when this assumption holds, RPSFTM performs pretty well[9,10]. However, this assumption may not hold if switching is only permitted after disease progression. It may be expected that the capacity for benefit amongst switchers is less than the one in patients initially randomized to the experimental group, especially when patients in control arm switch to multiple treatments with different levels of effects.

### 2.2.3 Iterative parameter estimation (IPE)

Branson and Whitehead proposed IPE method in 2002 as an extension of the RPSFTM method[11]. In general IPE method is similar to the RPSFTM method in the counterfactual framework except for estimating the acceleration factor by a parametric likelihood approach rather than the G-estimation. Therefore, it requires an additional assumption on the distributional form of the counterfactual survival time. With this method, a parametric failure time model is first fitted to the original, unadjusted ITT data to obtain an initial estimate of ψ. The observed failure time of switching patients are then re-estimated using the counterfactual survival time model, and the treatment groups are then compared again using a parametric failure time model. This gives an updated estimate of ψ and the process of re-estimating the observed survival time of switching patients is repeated. This iterative process continues until the new estimate for exp ($\psi$) converges.

## 3 Proposed Methods and Extension

In real clinical trials, it is possible that switchers switch to multiple treatments that differ from the experimental arm, where the assumption of "common treatment effect" for RPSFTM and IPE methods generally does not hold. In this section we propose a stratified RPSFTM as an extension to the original RPSFTM. The new method does not rely on the "common treatment effect" assumption and hence is suitable for settings where patients may switch to multiple levels.

In addition, we propose a random-forest-based prediction approach to directly predict a patient's counterfactual overall survival time $U_i$. This method can deal with even more treatment levels that switchers switch to and take care of the cases of the adjustment for switchers randomized to treatment arm.

### 3.1 Stratified RPSFTM

The first proposed method is the stratified RPSFTM, which is an extension based on RPSFTM, allowing multiple levels of treatment effect for switchers. In this method, patients in control arm are allowed to switch to multiple treatments. One of those treatments can be the experimental treatment or some equally efficacious treatments, and other treatments may not have the same effect as the experimental one. Therefore, instead of assuming "common treatment effect" across all switched arms, we restrict this assumption only for switchers that receive same or similar treatments as the experimental treatment. Switchers switching to other treatment may have different levels of treatment effect. In addition, similar to the original RPSFTM and IPE methods, it is still assumed that the counterfactual event time $U_i$ is independent of randomization and switching.

Similar to RPSFTM, the accelerated failure time (AFT) model is applied to model the counter factual event time $U_i$. First we split the observed event time $T_i$ for each patient into two parts: the event time when the patient is on the control treatment $T_{i,CON}$, and the event time when the patient is on the intervention treatment, $T_{i,TRT}$.

$$T_i = T_{i,CON} + T_{i,TRT}$$

Secondly, for patient *i* randomized to treatment arm or switchers switching to the same arm or treatment with similar effect as the experimental drug, we model the corresponding counterfactual event time $U_i$ with AFT model

$$U_i = T_{i,CON} + \exp(\psi_0) T_{i,TRT}$$

where $\exp(-\psi_0)$ is the acceleration factor corresponding to the experimental treatment. $\psi_0$ is the parameter to be estimated in this equation. If $\exp(\psi_0) < 1$ (equivalently $\psi_0 < 0$), the experimental treatment is beneficial.

Different parameters of $\psi$ are used in AFT model to model different levels of treatment effect. Assume that patient $i$ receive treatment level $k$ ($k = 1, ..., K$), we model the corresponding counterfactual event time $U_i$ with causal models

$$U_i = T_{i,CON} + \exp(\psi_k) T_{i,TRT}$$

where $\exp(-\psi_k)$ is the acceleration factor corresponding to the $k$-th level of treatment effect. If $\exp(-\psi_k) < \exp(-\psi_0)$ or equivalently $\psi_k > \psi_0$, the corresponding treatment is less effective compared with the experimental treatment and vice versa.

The parameter estimation method is motivated from the g-estimation of RPSFTM. Instead of estimating a single parameter $\psi$, we now estimate a series of parameters $\{\psi_k\}$, $k =0, ..., K$. Hence, we consider grids for $\{\psi_k\}$, $k =0, ..., K$, and test whether the distribution of $U_i$ is the same for all patients. For each combination of candidate values of $\{\psi_k\}$, $k =0, ..., K$, calculate $U_i$ and test the null hypothesis that $U_i$ is the same for all patients using a standard survival analysis test statistic Z. Finally, combination of $\{\psi_k\}$ that best satisfies test statistic Z = 0 is selected. Note that in this step, as in RPSFTM, re-censoring can be applied if needed.

After g-estimator for $\{\psi_k\}$, $k =0, ..., K$ are obtained, we can compute estimated counterfactual survival time for switchers, and perform ITT analysis using the observed $T_i$ for patients randomized to treatment arm and non-switchers in control arm, and the estimated counterfactual survival time $U_i$ for switchers.

There are two assumptions for stratified RPSFTM: the treatment effect of switchers can be categorized into several levels and $U_i$ distribution is the same for all patients. The first assumption is a relaxation of the "common treatment effect assumption" in RPSFTM allowing the use of several AFT models to estimate the counterfactual survival time. The second assumption is the same as RPSFTM allowing the model to obtain the g-estimation of parameters.

The strength of the stratified RPSFTM is obvious because it doesn't require the stronger assumption as compared to RPSFTM. Instead of assuming common treatment effect, it allows a few different levels of treatment effect, imitating more real-world scenarios. In addition, the estimation of counterfactual survival time follows AFT model and uses information both before and after switching. Hence the method is relatively robust to different proportions of switching comparing with IPCW method. However, the stratified RPSFTM may not be suitable if there are too many different levels of treatment effect to be estimated. Since each level of treatment effect is corresponding to one parameter, increasing number of levels add extra dimension of parameters to be estimated in the g-estimation step and increase computation burden.

## 3.2 Random Forest based prediction

IPCW methods mainly uses covariates to decide how important a patient is for the overall survival analysis in the presence of treatment switching, while RPSFTM models the counterfactual event time for switchers. In order to combine potential benefit of these two tracks, we propose a new method to predict counterfactual event time for switchers using covariates. To achieve nice prediction accuracy, random forest method is applied as our prediction model given its power to handle non-linearity and interaction of covariates.

### 3.2.1 Introduction to Random Forest method

Random forest[12] is a decision-tree based, easy to use machine learning algorithm that possesses strong prediction power of the response variable using covariates without any model form assumptions.

In general, a random forest regression algorithm is as follows:
- A bootstrap sample is drawn from the training data.
- Grow a tree to the bootstrapped data by selecting the best covariate and split-point in each splitting step such that the corresponding tree has the best prediction accuracy.
- Repeat the above two steps until a desired number of tree models is achieved.
- For any new data point, the final predicted value of response variable is the average of predicted value in all built trees.

Similar to decision tree, random forest is a non-parametric method with no assumption on model structures, which implies its flexibility in describing the unknown relationship between response variable and covariates compared with structured models, e.g. Cox PH model or AFT model. As a result, tree-based methods including random forest are especially valuable when there are many covariates and interactions need to be considered, as those models select important variables and relation structures automatically to achieve the best prediction accuracy.

Random forest is shown to improve the prediction accuracy of the tree method. Random forest predictors often have smaller variance as the prediction is obtained by averaging the result from several trees. In addition, random forest method uses thousands of bootstrap samples to build trees. It is less likely to overfit the data as compared with other prediction methods.

### 3.2.2 Random Forest based prediction approach in treatment switching

In this section we propose a new method to correct the bias of evaluating overall survival in the presence of treatment switching by predicting the counterfactual survival time $U_i$ using random forest. The method and algorithm is described below:

1) The whole sample is divided into training set and prediction set. The training set includes all the non-switchers in the control arm, with whom the observed survival time is exactly their counterfactual survival time. A random forest model to predict $U_i$ can be trained using the training set with all covariates.
2) The trained random forest model is then used predict the counterfactual event time for switchers. After prediction, the predicted counterfactual event time should be no shorter than the time of switching. To ensure this, the maximum of predicted value and time of switching is used as the final counterfactual survival time $U_i$.
3) Finally, ITT analysis is performed with observed survival time $T_i$ of non-switchers and predicted counterfactual survival time $U_i$ of switchers.

Unlike IPCW and RPSFTM, the assumptions for the random-forest-based prediction method rely on the framework of causal inference. Therefore, instead of assuming no unmeasured confounders or common treatment effect, this method assumes that the data include all important covariates for predicting counterfactual survival time. Furthermore, it is assumed that for patients in each arm, the counterfactual survival time can be well predicted using the same model regardless of the patients are censored or not.

The random-forest-based prediction method does not make assumption that patients can only switch to one single treatment, therefore this method allows patients to switch to any different treatments. It also deals with cases where patients in experimental arm also switch to other treatments. Furthermore, this method does not need any assumptions on model structures before fitting and can deal with large number of covariates, especially when there is no clear relationship between covariates and counterfactual survival time.

## 4  Simulation Study

In this section, the main objective is to evaluate existing methods and the newly proposed ones in the analysis of overall survival benefit of a treatment in the presence of treatment switching.

The setting of our simulation is designed based on a real Phase III study to investigate the efficacy and safety of a certain treatment in patients with a type of blood cancer. Approximately 400 subjects were randomly assigned in 1:1 ratio to receive either treatment arm or control arm. There are approximately 20% of events. About 50% of patients in the control arm switched to other treatments, with about 25% of patients switching to one particular treatment that is proved to have similar overall survival benefit as the experimental treatment. Patient's baseline age, baseline ECOG, number of previous treatment lines and risk levels are included as covariates.

In order to investigate the estimation accuracy of the newly proposed methods targeting on multiple switching scenarios, overall survival data are simulated for patients randomized to treatment arm and control arm, where the true hazard ratio of two arms are known. Patients in the control arm are allowed to switch to multiple treatments with different levels of effects. Both current and newly proposed treatment switching adjustment methods are applied to adjust for switching biases. For each scenario, 500 repetitions are run to get a reliable result.

## 4.1 Simulation scenarios

In this simulation, 400 patients are simulated each time as the sample size of the experiment. Three levels of HR are simulated: 0.4, 0.6 and 0.8 corresponding to large, medium or small treatment effect, respectively. As for proportion of censoring, 25%, 50% and 75% are used to represent low, moderate and high censoring rate of data. Similarly, the switch proportion within control arm is also considered in three levels: low (approximately 25% of control arm), moderate (approximately 50% of control arm) and high (approximately 75% of control arm).

The treatment effects for patients in the control arm switched to other treatments are divided into two levels. The first level is the experimental treatment effect, which can be considered as the effect of the experimental drug or equivalent ones; the second level is 30% less effective than the experimental effect, which can be considered as the effect of the secondary treatments such as standard of care.

## 4.2 Simulating survival time

First, patients are randomized to treatment arm and control arm with a ratio of 1:1. Their baseline age and ECOG are generated using Normal distributions to have similar quartiles as the real data; risk levels is a categorical variable generated from multinomial distribution. The covariates are generated to be mutually independent. The underlying survival time for treatment arm and control arm is generated using Weibull accelerated failure time (AFT) model

$$\log(S) = \alpha_0 + \alpha_1 \text{ arm} + \alpha_2 \text{baseline age} + \alpha_3 \text{baseline ECOG} + \alpha_4 \text{risk level} + \epsilon$$

$\alpha_1 = -\psi = -\log(\text{HR})$ controls the treatment effect. HR is one of the preset true hazard ratios: 0.4, 0.6 or 0.8. $\alpha_2 = -0.002, \alpha_3 = \alpha_4 = -0.05$ are chosen based on model estimation in the real study.

Second, the time of switching is generated from a Uniform distribution in (0, 365). Whether to switch or not is determined by a Bernoulli distribution with given switch proportion. For switchers the new survival time after switching is computed using AFT model:

$$T_{i,TRT} = T_{i,CON(\text{after switching})} \times \exp(-\psi_k) \times F$$

where the choice of F is one of the candidates 1.0 and 0.7, representing the experimental treatment effect and the secondary treatment effect which is 70% of the experimental one, respectively. The final observed survival time for switchers are

$$T_i = T_{i,CON\text{(before switch)}} + T_{i,TRT}$$

## 4.3 Results

For each simulated dataset, analysis is conducted using the following seven methods: ITT, RPSFTM, IPE, censoring switchers, IPCW, random forest (RF) and Stratified RPSFTM (SRP). The performance of each method differs depending upon the scenarios investigated.

### 4.3.1 Estimated HR and 95% CI

The performance of each method in estimating hazard ratio is summarized in Figure 1 where true hazard ratio equals 0.4, 0.6 and 0.8 respectively. In each case, HR and 95% CI are estimated using the seven methods for different combination of censoring rates and switch rates based on 500 repetitions.

According to Figure 1, 2 and 3, it can be seen that in most cases all methods tend to overestimate the true HR, or equivalently, underestimate the overall survival benefit for experimental treatment. For HR = 0.4 and 0.6, biases of estimations for all methods increase as censoring rate and switch rate get larger. When HR = 0.8, the treatment effect is quite small, all methods performs closely except IPCW and censoring switcher methods. More specifically, IPCW and censoring switchers become considerably biased and unstable when both censoring and switch rates are high. In addition, for large censoring rates, stratified RPSFTM(SRP) usually yields smaller biases and narrower CIs while for small censoring rate (25%) and medium to high switch rate (50% to 75%), SRP tends to underestimate the true HR. In this scenario, random forest method gives less biased and relatively robust estimators.

### 4.3.2 Metrics for Evaluation

Throughout our simulation, three metrics are applied for evaluation of treatment switching adjustment methods:
1. Bias = true HR - average estimated HR
2. Mean squared error (MSE) = mean[estimated HR − true HR]$^2$
3. Coverage ratio: The proportion of CI that covers the true HR

These three numerical metrics are summarized for the seven evaluated methods in table 1, 2 and 3, respectively, for all scenarios. The best result among the seven methods in each scenario is highlighted in red. According to these tables, the current existing methods tend to over-estimate HR, especially when censoring and switch rates are high. In general, RPSFTM and IPE perform the best among all current methods. ITT gives relatively large estimated HR values, but has decent performance when true HR increases to 0.8. Censoring and IPCW are really sensitive to high censoring and switch rates and are only reliable for switch rates no larger than approximately 0.6. In terms of coverage probability, IPCW and censoring switchers methods are least recommended.

In all scenarios our newly proposed methods show benefit over current existing methods among the three numerical metrics from the simulation. Stratified RPSFTM outperforms all methods in most scenarios except that it underestimates HR when censoring rate is as small as 0.25, in which case random forest is much more stable. Recommendations of methods under different censoring rate, switch proportion and HR are summarized in Table 4.

# 5   Conclusion and Discussion

In this paper, we summarized popular existing adjustment methods for overall survival analysis in the presence of treatment switching. Because the existing methods require strong assumptions that may not be satisfied in real world setting and are limited in the setting of switching to single treatment, we proposed two new methods to relax the strong assumption and obtain more robust estimator, especially in the general setting that patients in control arm switched to multiple treatments.

The first method is the stratified RPSFTM that is an extension to RPSFTM from single treatment switching to multiple treatment switching. For each additional level of treatment, an additional AFT model is used to describe the new treatment effect. Hence the method has the assumption that the counterfactual survival time is the same for every patient. This method can deal with cases where patients in control arm switch to a few different levels of treatment. The estimation of stratified RPSFTM still follows the g-estimation structure and is similar to the parameter estimation procedure in Robins, J. M., & Greenland, S. (1994)[13], with multiple acceleration parameters to estimate. The second method is to predict counterfactual event time using random forest model. Random forest method is one of the most reliable prediction models and the proposed method takes covariates into consideration without breaking the randomization. It has much relaxed assumption compared to IPCW and provides more stable and less biased result when switch rate is high.

In our simulation, we evaluate existing and newly proposed methods in different scenarios in the presence of multiple switching. Simulation results show that when HR is as high as 0.8, switching or not often does not affect the overall survival benefit too much and the ITT method performs well. In other scenarios, random-forest-based method provides better result when censoring rate is low, as in those cases the relatively large training sample sizes are available to train reliable random forest models. Otherwise, stratified RPSFTM outperforms all other methods.

When switch rate is at moderate to high level, IPCW and censoring methods tend to have large estimation biases and wide confidence intervals. This shows the major drawback of IPCW: when proportion of switch is high, the IPC weights for non-switchers become very large, which causes large biases and instability in estimation. The similarity of the IPCW and censoring methods is due to the fact that, when covariates have much smaller contribution to overall survival compared with treatment effect, IPC weights are closed to 1 and IPCW methods is not very effective in correcting the switching biases.

In addition, RPSFTM performs similarly to stratified RPSFTM. This is probably due to the fact that the two methods share the same AFT model except for using different acceleration parameters. In our simulation, the two new switched? treatments have 70% of experimental treatment effect. If the number of new treatment increases, or the new treatment effect differs a lot from the experimental treatment, it is reasonable to expect a larger difference in estimation of HR using RPSFTM and stratified RPSFTM.

Even though the newly proposed methods improve and extend current methods in different directions, there are still limitations for each of the methods. For stratified RPSFTM, the current parameter estimation method is based on the g-estimation in RPSFTM. However, as the number of new treatment levels increases, the computation burden of g-estimation increases dramatically because the g-estimation is based on grid search of the best candidate of parameter values. Hence the stratified RPSFTM is more suitable for small number of new treatment levels (usually fewer than four levels). If better parameter estimation method or algorithm is developed in the future, this method can be less limited. For random-forest-based method, it is required that for all patients in each arm, regardless of censored or dead, the counterfactual survival time can be well predicted using the same model. This is a relatively strong assumption and hard to check. Future improvement of this method can be targeted on relaxing the strong assumptions.

In summary, the two new proposed methods should add to the current literature for analyzing overall survival with treatment switching and fill the methodologic gaps that still exist in the field.

# 7 Tables

Table 1 MSE ($\times 10^2$) for seven evaluated methods in all scenarios.

| Methods | MSE | HR = 0.4 | | | HR = 0.6 | | | HR = 0.8 | | |
|---|---|---|---|---|---|---|---|---|---|---|
| | | 25% Censored | 50% Censored | 75% Censored | 25% Censored | 50% Censored | 75% Censored | 25% Censored | 50% Censored | 75% Censored |
| ITT | 25% Switched | 0.826 | 0.853 | 1.236 | 0.753 | 1.023 | 1.953 | 0.888 | 1.364 | 2.734 |
| RPSFTM | | 0.290 | 0.445 | 0.96 | 0.612 | 0.929 | 1.961 | 1.096 | 1.634 | 3.130 |
| IPE | | 0.305 | 0.462 | 0.979 | 0.609 | 0.928 | 1.959 | 1.074 | 1.609 | 3.094 |
| Censoring | | 0.510 | 0.832 | 1.714 | 1.234 | 2.031 | 4.213 | 2.372 | 3.968 | 7.990 |
| IPCW | | 0.516 | 0.838 | 1.727 | 1.245 | 2.046 | 4.232 | 2.392 | 3.985 | 8.042 |
| RF | | 0.294 | 0.501 | 1.102 | 0.583 | 0.915 | 1.946 | 0.989 | 1.466 | 2.849 |
| SRP | | 0.275 | 0.394 | 0.953 | 0.594 | 0.875 | 1.885 | 1.059 | 1.528 | 2.922 |
| ITT | 50% Switched | 2.543 | 2.367 | 2.799 | 1.405 | 1.659 | 2.703 | 0.929 | 1.457 | 2.665 |
| RPSFTM | | 0.391 | 0.604 | 1.551 | 0.856 | 1.229 | 2.585 | 1.585 | 2.212 | 3.600 |
| IPE | | 0.499 | 0.760 | 1.720 | 0.815 | 1.225 | 2.586 | 1.413 | 2.040 | 3.470 |
| Censoring | | 1.964 | 3.414 | 7.679 | 5.499 | 9.708 | 20.50 | 11.45 | 20.16 | 40.63 |
| IPCW | | 1.982 | 3.440 | 7.721 | 5.542 | 9.733 | 20.61 | 11.52 | 20.20 | 40.84 |
| RF | | 0.357 | 0.732 | 1.967 | 0.711 | 1.113 | 2.517 | 1.169 | 1.631 | 2.891 |
| SRP | | 0.456 | 0.467 | 1.451 | 0.883 | 1.054 | 2.365 | 1.413 | 1.782 | 2.990 |
| ITT | 75% Switched | 5.768 | 4.876 | 4.876 | 2.185 | 2.431 | 3.085 | 0.916 | 1.572 | 2.250 |
| RPSFTM | | 0.591 | 0.758 | 0.758 | 1.436 | 1.627 | 2.815 | 2.645 | 3.258 | 3.686 |
| IPE | | 1.124 | 1.330 | 1.330 | 1.079 | 1.486 | 2.789 | 1.876 | 2.717 | 3.367 |
| Censoring | | 12.313 | 19.989 | 19.989 | 35.537 | 62.642 | 153.72 | 79.035 | 145.53 | 320.67 |
| IPCW | | 12.528 | 20.377 | 20.377 | 36.142 | 64.727 | 152.16 | 80.128 | 151.99 | 317.85 |
| RF | | 0.488 | 0.860 | 0.860 | 0.910 | 1.274 | 2.719 | 1.393 | 1.950 | 2.550 |
| SRP | | 0.988 | 0.597 | 0.597 | 1.586 | 1.260 | 2.194 | 2.017 | 2.256 | 2.706 |

Table 2 Bias($\times 10^2$) for seven evaluated methods in all scenarios.

| Methods | Bias | HR = 0.4 | | | HR = 0.6 | | | HR = 0.8 | | |
|---|---|---|---|---|---|---|---|---|---|---|
| | | 25% Censored | 50% Censored | 75% Censored | 25% Censored | 50% Censored | 75% Censored | 25% Censored | 50% Censored | 75% Censored |
| ITT | 25% Switched | -7.225 | -6.387 | -5.559 | -4.569 | -4.245 | -4.159 | -1.443 | -1.484 | -1.829 |
| RPSFTM | | -0.245 | -1.318 | -2.386 | 0.257 | -0.486 | -1.763 | 1.142 | 0.561 | -0.651 |
| IPE | | -0.855 | -1.732 | -2.619 | -0.181 | -0.833 | -1.964 | 0.929 | 0.365 | -0.761 |
| Censoring | | -4.395 | -5.896 | -7.838 | -7.445 | -9.638 | -12.71 | -10.68 | -13.80 | -18.46 |
| IPCW | | -4.420 | -5.916 | -7.862 | -7.476 | -9.664 | -12.72 | -10.71 | -13.81 | -18.49 |
| RF | | -0.920 | -2.807 | -4.279 | -0.913 | -2.252 | -3.601 | -0.530 | -1.165 | -2.181 |
| SRP | | 1.381 | -0.554 | -2.543 | 1.234 | -0.532 | -2.204 | 0.888 | -0.026 | -1.276 |
| ITT | 50% Switched | -14.64 | -13.39 | -12.48 | -8.887 | -8.435 | -8.236 | -2.320 | -2.531 | -1.992 |
| RPSFTM | | 0.422 | -2.005 | -5.371 | 1.646 | -0.270 | -3.234 | 3.650 | 2.026 | 0.613 |
| IPE | | -2.570 | -4.051 | -6.584 | -0.453 | -1.823 | -4.126 | 2.547 | 1.136 | 0.116 |
| Censoring | | -11.910 | -15.83 | -22.89 | -20.54 | -27.16 | -37.48 | -30.03 | -39.55 | -53.74 |
| IPCW | | -11.968 | -15.88 | -22.90 | -20.62 | -27.20 | -37.52 | -30.13 | -39.58 | -53.79 |
| RF | | -0.612 | -4.417 | -8.787 | -0.404 | -3.195 | -6.514 | 0.577 | -0.949 | -2.325 |
| SRP | | 4.037 | -0.085 | -5.599 | 3.870 | 0.012 | -4.129 | 3.444 | 1.209 | -0.456 |
| ITT | 75% Switched | 5.768 | 4.876 | 4.876 | -12.27 | -11.88 | -10.84 | -2.579 | -2.605 | -2.306 |
| RPSFTM | | 0.591 | 0.758 | 0.758 | 5.331 | 1.515 | -2.514 | 8.317 | 5.244 | 2.608 |
| IPE | | 1.124 | 1.330 | 1.330 | -0.345 | -2.334 | -4.978 | 5.055 | 3.003 | 1.059 |
| Censoring | | 12.313 | 19.989 | 19.989 | -56.52 | -73.96 | -112.14 | -84.51 | -112.5 | -162.43 |
| IPCW | | 12.528 | 20.377 | 20.377 | -56.92 | -75.08 | -111.81 | -84.80 | -114.4 | -161.59 |
| RF | | 0.488 | 0.860 | 0.860 | 1.452 | -3.239 | -8.361 | 2.496 | -0.259 | -2.845 |
| SRP | | 0.988 | 0.597 | 0.597 | 7.946 | 0.924 | -4.231 | 7.165 | 2.900 | 1.399 |

Table 3 Coverage percentage for seven evaluated methods in all scenarios.

| Methods | Coverage | HR = 0.4 | | | HR = 0.6 | | | HR = 0.8 | | |
|---|---|---|---|---|---|---|---|---|---|---|
| | | 25% Censored | 50% Censored | 75% Censored | 25% Censored | 50% Censored | 75% Censored | 25% Censored | 50% Censored | 75% Censored |
| ITT | 25% Switched | 74.6% | 84.4% | 91.4% | 91.0% | 92.8% | 93.2% | 95.0% | 94.6% | 94.8% |
| RPSFTM | | 95.8% | 95.0% | 94.4% | 93.6% | 94.6% | 93.8% | 92.8% | 92.6% | 93.2% |
| IPE | | 95.0% | 93.8% | 94.4% | 93.4% | 94.2% | 93.8% | 93.0% | 92.8% | 93.2% |
| Censoring | | 87.2% | 85.6% | 85.6% | 80.2% | 81.2% | 83.0% | 76.4% | 76.2% | 76.2% |
| IPCW | | 87.2% | 85.8% | 85.4% | 80.4% | 81.6% | 82.8% | 76.2% | 76.4% | 76.4% |
| RF | | 94.4% | 93.4% | 93.4% | 94.8% | 94.0% | 93.0% | 93.4% | 93.6% | 94.0% |
| SRP | | 96.6% | 96.6% | 94.8% | 94.2% | 95.2% | 93.4% | 92.2% | 93.8% | 93.8% |
| ITT | 50% Switched | 37.0% | 59.2% | 81.0% | 81.0% | 87.4% | 91.2% | 94.8% | 95.0% | 94.6% |
| RPSFTM | | 96.0% | 94.4% | 92.6% | 91.8% | 93.4% | 92.4% | 86.6% | 88.8% | 91.4% |
| IPE | | 91.8% | 91.4% | 90.8% | 92.2% | 92.8% | 92.6% | 88.8% | 90.2% | 91.6% |
| Censoring | | 56.0% | 50.8% | 53.4% | 33.4% | 33.0% | 38.2% | 22.8% | 22.0% | 27.2% |
| IPCW | | 54.6% | 49.2% | 52.8% | 34.2% | 32.6% | 37.8% | 20.4% | 20.8% | 27.0% |
| RF | | 96.0% | 90.6% | 87.8% | 93.8% | 93.2% | 91.6% | 91.8% | 94.0% | 92.4% |
| SRP | | 95.8% | 97.6% | 93.0% | 91.4% | 94.4% | 92.8% | 87.6% | 92.0% | 93.0% |
| ITT | 75% Switched | 3.2% | 15.6% | 50.0% | 64.8% | 79.6% | 87.2% | 95.0% | 94.6% | 95.6% |
| RPSFTM | | 77.4% | 84.8% | 81.2% | 72.4% | 82.2% | 87.4% | 66.8% | 77.6% | 86.0% |
| IPE | | 68.4% | 70.8% | 74.4% | 81.8% | 85.6% | 88.6% | 78% | 83.8% | 88.6% |
| Censoring | | 6.2% | 5.4% | 6.3% | 2.4% | 2.4% | 3.6% | 0.4% | 0.8% | 3.6% |
| IPCW | | 7.0% | 5.6% | 6.4% | 2.4% | 2.4% | 3.6% | 0.6% | 0.6% | 3.2% |
| RF | | 84.4% | 82.8% | 70.4% | 84.6% | 90.4% | 88.6% | 84.0% | 90.6% | 93.0% |
| SRP | | 54.4% | 87.4% | 80.6% | 68.2% | 87.4% | 89.8% | 75.2% | 85.4% | 90.6% |

Table 4 Recommendations of methods under different censor rate, switch proportion and expected HR

| Recommended Method | | 25% Censored | 50% Censored | 75% Censored |
|---|---|---|---|---|
| **HR = 0.4** | 25% Switched | SRP | SRP | SRP |
| | 50% Switched | RF | SRP | SRP |
| | 75% Switched | RF | SRP | SRP |
| **HR = 0.6** | 25% Switched | RF | SRP | SRP |
| | 50% Switched | RF | SRP | SRP |
| | 75% Switched | RF | SRP | SRP |
| **HR = 0.8** | 25% Switched | ITT | ITT | ITT |
| | 50% Switched | ITT | ITT | ITT |
| | 75% Switched | ITT | ITT | ITT |

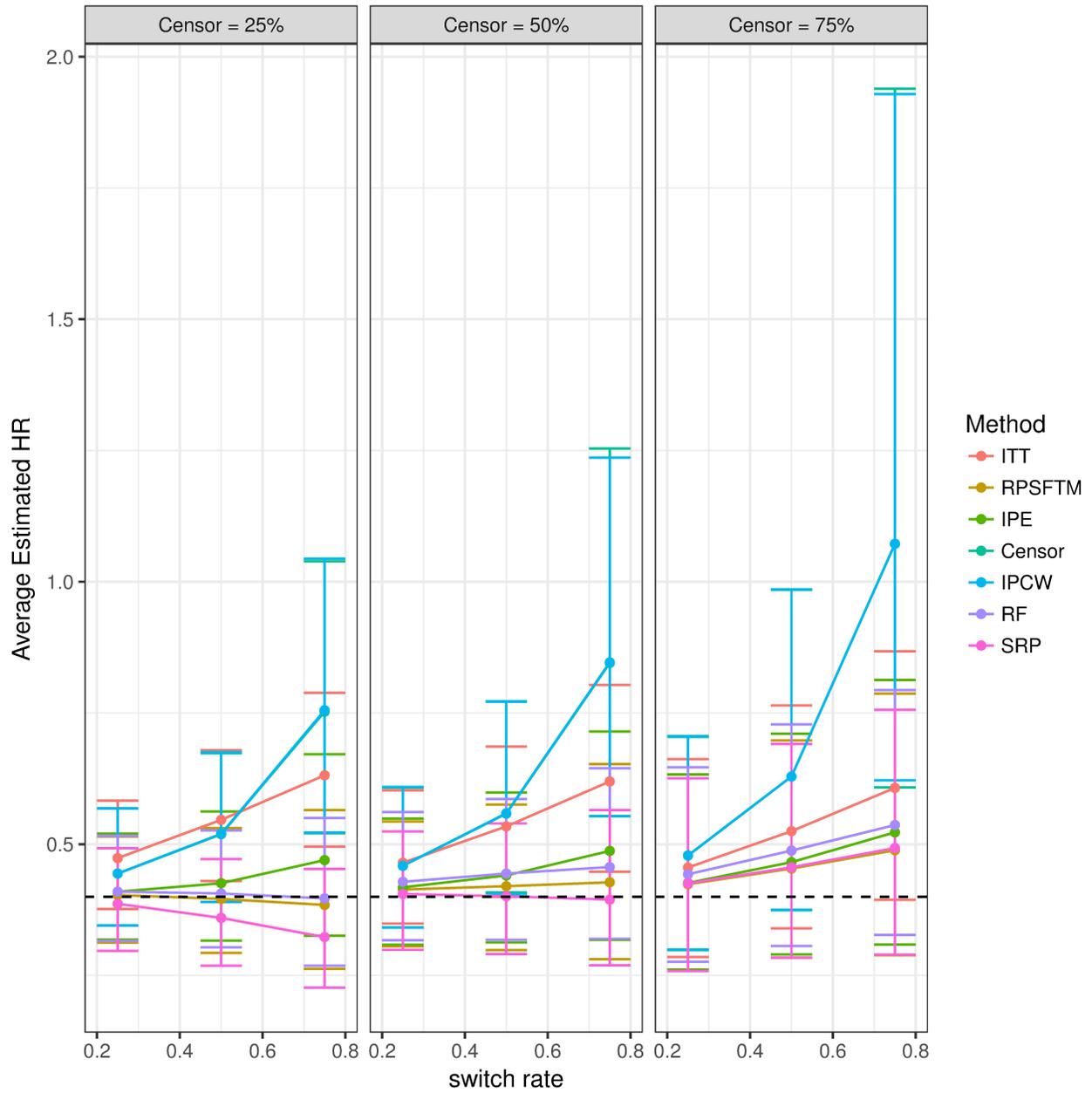

Figure 1: Estimated HR for seven evaluated methods under scenarios when true HR = 0.4.

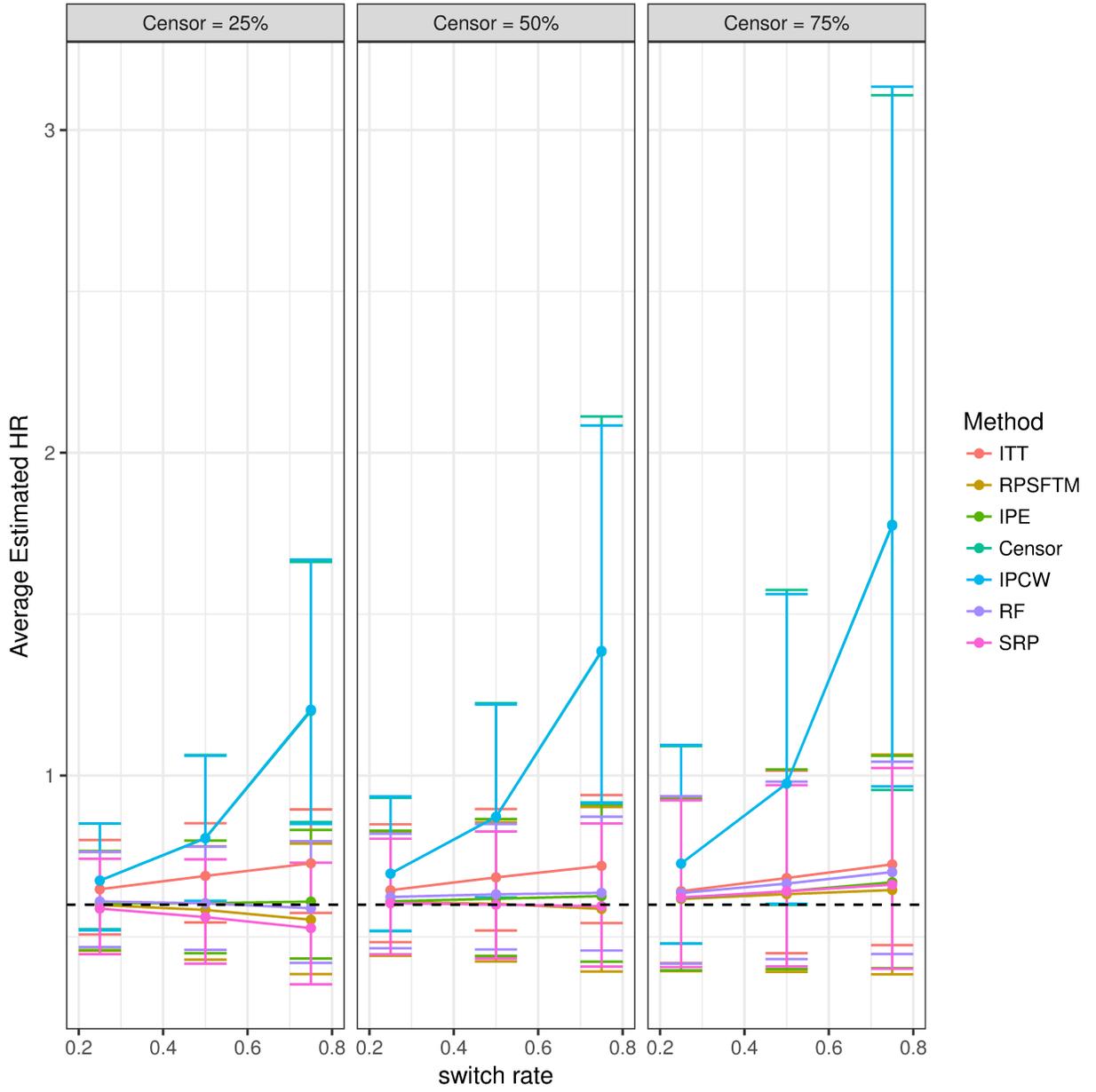

Figure 2: Estimated HR for seven evaluated methods under scenarios when true HR = 0.6

Figure 3: Estimated HR for seven evaluated methods under scenarios when true HR = 0.8

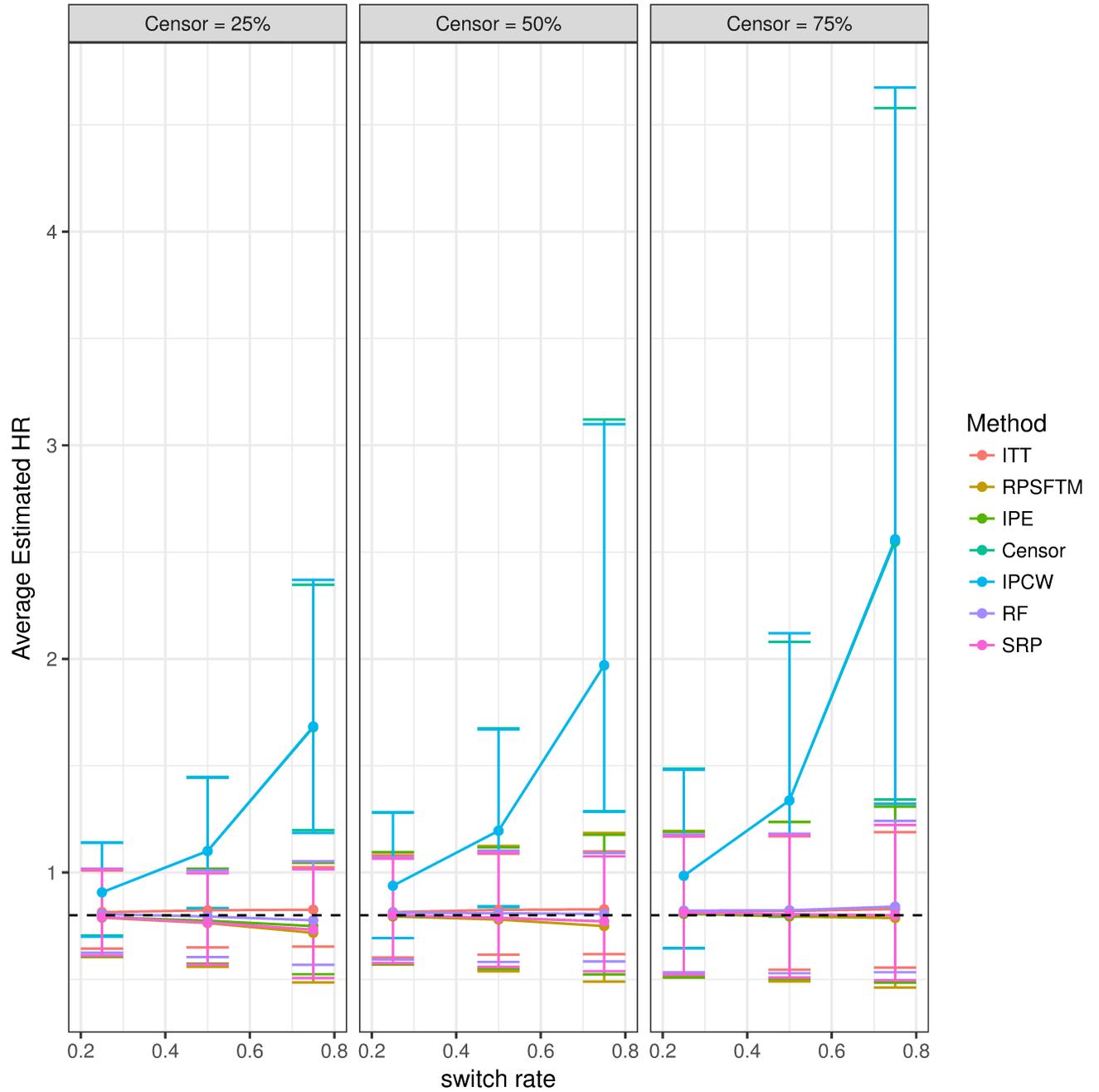